\documentclass[conference]{IEEEtran}

\usepackage{textcomp}

\usepackage{amsmath}
\usepackage{amssymb}
\usepackage{graphicx}
\usepackage{tikz}
\usepackage{multicol}
\usepackage{pifont}
\usepackage{enumerate}

\newtheorem{mylemma}{\it Lemma}
\newtheorem{mytheorem}{\it Theorem}
\newtheorem{myproposition}{\it Proposition}
\newtheorem{myproof}{\it Proof}
\newtheorem{mydefinition}{\it Definition}
\newtheorem{myexample}{\it Example}

\newcommand{\MYfooter}{\smash{
\hfil\parbox[t][\height][t]{\textwidth}{}\hfil\hbox{}}}
\makeatletter
\def\@oddhead{\mbox{}2018 ICSEE International Conference on the Science of Electrical Engineering \rightmark \hfil }%
\def\@oddfoot{\MYfooter}%
\def\ps@IEEEtitlepagestyle{
  \def\@oddfoot{\mycopyrightnotice}
  \def\@evenfoot{}
}
\def\mycopyrightnotice{
}

\begin{document}

\title{Non Binary Polar Codes with Equidistant Transform for Transmission over the AWGN Channel}

\author{\IEEEauthorblockN{Sinan Kahraman}
\IEEEauthorblockA{Department of Electrical-Electronics Engineering\\
Bilkent University\\
Ankara, TR-06800, Turkey\\
Email: sinank@ieee.org}
}

\maketitle

\begin{abstract}
Finding fast polarizing transforms is an important problem as polar codes suffer from slow finite­length performance. This paper considers non­ binary polar codes for transmission over the AWGN channel and designs polarizing transforms with better distance characteristics using a simple procedure for signal sets. The main idea of the paper is to define "Equidistant Polarizing Transforms" and show that they achieve the best distance spectrum bound. 
We provides an example of such transform for $q = 5$. In this case PSK-type signal set is used. 
Finally, we show performance gains and some comparison with other methods. 
\end{abstract}

\section{Introduction} 
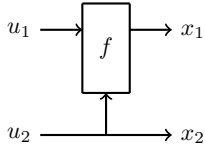
\begin{figure}[hp]
\centering
\begin{tikzpicture}[thick,scale=0.7, every node/.style={scale=0.9}]
\draw[->]  (-1,0)  -- (-0.2,0) ;
\draw[->]  (0.7,0)  -- (1+1-0.5,0) ;
\draw[->]  (0.25,-2)  -- (0.25,-1.2) ;
%\draw  (-0.2,-1-0.2)  -- (0.2,-1-0.2) ;
%\draw  (-0.2,-1+0.2)  -- (0.2,-1+0.2) ;
%\draw  (-0.2,-1-0.2)  -- (-0.2,-1+0.2) ;
%\draw  (0.2,-1-0.2)  -- (0.2,-1+0.2) ;
%\draw  (0,-0.2)  -- (0,0.2) ;
\draw[->]  (-1,-2)  -- (1+1-0.5,-2) ;
\draw (-0.2,-1.2)  -- (0.7,-1.2);
\draw (-0.2,0.5)  -- (0.7,0.5);
\draw (-0.2,-1.2)  -- (-0.2,0.5);
\draw (0.7,-1.2)  -- (0.7,0.5);
\begin{scope}[thick]
%\draw (0,0) circle (0.2cm);
\draw (0.25,-0.35) node[] {$f$};
\end{scope}
%\draw (1.9,0.8) node[left] {\em mod-q adder};
%\draw[ thin] [->] (0.15,0.45)  -- (0.05,0.25);
%\draw[ thin]       (0.15,0.65)  -- (0.05,0.45);
%\draw[ thin]       (0.05,0.45)  -- (0.15,0.45);
%\draw (0.8,-1) node[right] {\em permutation};
%\draw[ thin] [->] (0.65,0.45-1.35)  -- (0.45,0.35-1.35);
%\draw[ thin]       (0.85,0.45-1.35)  -- (0.65,0.35-1.35);
%\draw[ thin]       (0.65,0.35-1.35)  -- (0.65,0.45-1.35);
\draw (-1,0) node[left] {$u_1$};
\draw (-1,-2) node[left] {$u_2$};
\draw (1+2-1.5,0-0.03) node[right] {$x_1$};
\draw (1+2-1.5,-2-0.03) node[right] {$x_2$};
\end{tikzpicture}
\caption{A basic scheme with a polarizing transform $f$ for q-ary input alphabet.}
\label{Fig_1}
\end{figure}

This paper presents a method for improving distance characteristics of non-binary polar codes. Following the notation in \cite{arikan_channel_2009}, we consider a memoryless channel $W:\mathcal{X} \rightarrow \mathcal{Y}$ with input alphabet $\mathcal{X}$, output alphabet $\mathcal{Y}$, and transition probabilities $\{W(y|x):x \in \mathcal{X} ,y \in \mathcal{Y} \}$. We assume that $\mathcal{X}$ is a finite set and label its elements such that $\mathcal{X}=\{0,1,\dots,q-1\}$, where $q\geq2$ is an arbitrary integer. We leave $\mathcal{Y}$ arbitrary for the moment. We will consider polarization schemes based on a basic transform of the type depicted in Fig.~\ref{Fig_1}.

The transform is defined by a \textsl{kernel} $$f:\mathcal{X}^2 \rightarrow \mathcal{X},$$ which we assume is a mapping with the following properties:
\begin{itemize}
\item for any fixed $u_1 \in \mathcal{X}, u_2 \rightarrow f(u_1,u_2)$ is an invertable function of $u_2$;
\item for any fixed $u_2 \in \mathcal{X}, u_1 \rightarrow f(u_1,u_2)$ is an invertable function of $u_1$. 
\end{itemize}
The standard polar coding kernel as defined in \cite{arikan_channel_2009} is a mapping of this type with $f(u_1,u_2)=u_1\oplus u_2$, where $\oplus$ denotes addition mod-$q$. The present paper shows that it is possible to construct polar codes with better distance properties (hence better performance) using alternative kernels of the above type.

\section{System Model}\label{S_system_model}
In this study, we first consider PSK signal sets for $q$-ary input alphabets. 
Let $\mathcal{S}$ be a signal set with size $q$, where $s_k=\sqrt{E_s}e^{2\pi k/q}\in \mathcal{S}$ for $k=0,1,\dots,q-1$. Here, the signal energy is $E_s$ joule/2-dimensions, and the $q$-ary PSK signal sets $\mathcal{S}:\left\{\sqrt{E_s},\sqrt{E_s} e^{j\frac{2\pi}{q}},\dots,\sqrt{E_s} e^{j\frac{2\pi(q-1)}{q}}\right\}$ for a given $q$.

Here, a natural mapping $x_i\rightarrow s_{x_i}$ is considered for transmission on AWGN channel. %+ 
The noisy observations from the channel are defined for $i=1,\dots,N$ as follows: 
\begin{equation}
{ y_i=s_{x_i}+n_i},
\end{equation} 
where $n_i$ is a complex Gaussian random variable with $\mathcal{CN}(0,\sigma^2)$. %+
The power spectral density is $\sigma^2=N_0$ joule/2-dimensions. %+
Hence, the signal to noise ratio is ${E_s}/{N_0}$. 

The transition probabilities of the synthetic channels obtained after one-step of polarization are defined as follows:
\begin{eqnarray}
W(y_1,y_2|u_1)=\frac{1}{q}\sum_{u_2=0}^{q-1}W(y_1|f(u_1,u_2))W(y_2|u_2),   \\
W(y_1,y_2,u_1|u_2)=\frac{1}{q}W(y_1|f(u_1,u_2))W(y_2|u_2),
\end{eqnarray}
where $W(y|x)=\frac{1}{\pi \sigma^2}e^{-{\|y-s_x\|^2}/{\sigma^2}}$. Here, $W(y_1,y_2,u_1|u_2)$ is known as polarized good channel. $W(y_1,y_2|u_1)$ is known as polarized bad channel. These transition probabilities describe a link between $f$ and error performances for polarized channels.  Note that $\sigma^2$ is the variance for 2-dimensions. Squared \textsl{Euclidean} distance is denoted by $\|\cdot\|^2$.

For the synthetic good channel, the distance for a given $u_1$ and $u_2\neq u_2'$ for a given transform $f$ is 
\begin{eqnarray}
d=\sqrt{\|s_{f(u_1,u_2)}-s_{f(u_1,u_2')}\|^2+\|s_{u_2}-s_{u_2'}\|^2}.
\end{eqnarray} 
The minimum distance for the standard transform is
\begin{eqnarray}
d_{min}=\sqrt{\|s_{m}-s_{m+1}\|^2+\|s_{n}-s_{n+1}\|^2}.
\end{eqnarray} 
The minimum distance for the PSK signal set is given by the following equation that is given for the standard transform.
\begin{eqnarray}
d_{min}&=&2\sqrt{2}\sin\left({\pi}/{q}\right)\sqrt{E_s}
\label{mindist}
\end{eqnarray}

In the next section, we will describe design of transforms which improve distance properties.

\section{Design of Polarizing Transforms}\label{S_design_polarization}
For the AWGN channel, the distance spectrum upper bound is given by symbol error probability $P_e$ as follows:
\begin{eqnarray}
P_e\leq\sum_{d\geq d_{min}} N(d)\cdot Q\left({d}/{(2\sigma)}\right),
\label{dsb}
\end{eqnarray}  
where $Q(d/(2\sigma))$ corresponds to the pairwise error probability between two points at a distance $d$ apart, and $N(d)$ denotes the distance spectrum defined as the number of points at distance $d$\footnote{Obviously, $Q(x)=\frac{1}{\sqrt{2\pi}}\int_{x}^{\infty} e^{-t^2/2} dt$  for the channel with $\sigma=1$. Notice that $\sigma^2$ is for 1-dimension in (\ref{dsb}), and $SNR=E_s/N_0$ where the signal power is $E_s$ joule/2-dimensions and $N_0$ is $\sigma^2$ joule/2-dimensions.}.

We investigate the distance properties by using a table where $u_1$ and $u_2$ are shown in a cell with coordinates $(x_1,x_2)$  corresponding to the outputs of the scheme in Fig.~\ref{Fig_1}. For more clarity, the cells are marked with a grey face for $u_1=0$.
We give a table for the standard transform (i.e. with the type of $f(u_1,u_2)=u_1\oplus \pi_0(u_2)$, where $\pi_0$ is the identical permutation) for $q=5$ as follows:
$$
\begin{tikzpicture}[scale=0.85, every node/.style={scale=0.85}]
\draw  (-1.5,0)  -- (1.5,0) ;%+++++
\draw  (-1.0,-0.5)  -- (1.5,-0.5) ;%+++++
\draw  (-1.0,-1.0)  -- (1.5,-1.0) ;%+++++
\draw  (-1.0,-1.5)  -- (1.5,-1.5) ;%+++++
\draw  (-1.0,-2.0)  -- (1.5,-2.0) ;%+++++
\draw  (-1.0,-2.5)  -- (1.5,-2.5) ;%+++++
\draw  (-1.0, 0.5)  -- (-1.0,-2.5) ;%+++++
\draw  (-0.5, 0.0)  -- (-0.5,-2.5) ;%+++++
\draw  (-0.0, 0.0)  -- (-0.0,-2.5) ;%+++++
\draw  (0.5, 0.0)  -- (0.5,-2.5) ;%+++++
\draw  (1.0, 0.0)  -- (1.0,-2.5) ;%+++++
\draw  (1.5, 0.0)  -- (1.5,-2.5) ;%+++++
\draw  (-1.5, 0.5)  -- (-1.0,0.0) ;%+++++
\draw (-1.2,0.5) node[] {$x_1$};
\draw (-0.75,0) node[above] {$0$};
\draw (-0.25,0) node[above] {$1$};
\draw ( 0.25,0) node[above] {$2$};
\draw ( 0.75,0) node[above] {$3$};
\draw ( 1.25,0) node[above] {$4$};
\draw (-1.5,-0.0) node[above] {$x_2$};
\draw (-1.25,-0.5) node[above] {$0$};
\draw (-1.25,-1.0) node[above] {$1$};
\draw (-1.25,-1.5) node[above] {$2$};
\draw (-1.25,-2.0) node[above] {$3$};
\draw (-1.25,-2.5) node[above] {$4$};
\draw[fill=gray!50] (-1.25+0.5-0.25,-0.5) rectangle (-1.25+0.5+0.25,-0.5+0.5) (-1.25+0.5,-0.5) node[above] {$00$};
\draw (-1.25+0.5,-1.0) node[above] {$41$};
\draw (-1.25+0.5,-1.5) node[above] {$32$};
\draw (-1.25+0.5,-2.0) node[above] {$23$};
\draw (-1.25+0.5,-2.5) node[above] {$14$};
\draw (-1.25+1.0,-0.5) node[above] {$10$};
\draw[fill=gray!50] (-1.25+1.0-0.25,-1.0) rectangle (-1.25+1.0+0.25,-1.0+0.5) (-1.25+1.0,-1.0) node[above] {$01$};
\draw (-1.25+1.0,-1.5) node[above] {$42$};
\draw (-1.25+1.0,-2.0) node[above] {$33$};
\draw (-1.25+1.0,-2.5) node[above] {$24$};
\draw (-1.25+1.5,-0.5) node[above] {$20$};
\draw (-1.25+1.5,-1.0) node[above] {$11$};
\draw[fill=gray!50] (-1.25+1.5-0.25,-1.5) rectangle (-1.25+1.5+0.25,-1.5+0.5) (-1.25+1.5,-1.5) node[above] {$02$};
\draw (-1.25+1.5,-2.0) node[above] {$43$};
\draw (-1.25+1.5,-2.5) node[above] {$34$};
\draw (-1.25+2.0,-0.5) node[above] {$30$};
\draw (-1.25+2.0,-1.0) node[above] {$21$};
\draw (-1.25+2.0,-1.5) node[above] {$12$};
\draw[fill=gray!50] (-1.25+2.0-0.25,-2.0) rectangle (-1.25+2.0+0.25,-2.0+0.5) (-1.25+2.0,-2.0) node[above] {$03$};
\draw (-1.25+2.0,-2.5) node[above] {$44$};
\draw (-1.25+2.5,-0.5) node[above] {$40$};
\draw (-1.25+2.5,-1.0) node[above] {$31$};
\draw (-1.25+2.5,-1.5) node[above] {$22$};
\draw (-1.25+2.5,-2.0) node[above] {$13$};
\draw[fill=gray!50] (-1.25+2.5-0.25,-2.5) rectangle (-1.25+2.5+0.25,-2.5+0.5) (-1.25+2.5,-2.5) node[above] {$04$};
\draw (0.25,-2.5) node[below] {$\pi_0=\left(\begin{smallmatrix} 0& 1& 2& 3& 4\\ 0&1&2&3&4 \end{smallmatrix}\right).$};
\end{tikzpicture}$$

The minimum distance of the standard transform is 
\begin{equation}
d_{\min}=\sqrt{2}\|s_0-s_1\|=1.66\sqrt{E_s}
\label{dmin_st}
\end{equation} 
for 5-PSK signalling.
The distance spectrum $N(d)$ for the standard transform is given as follows:   
\begin{equation}
N(d)=\left\{
\begin{array}{ll}
2   & d=1.66\sqrt{E_s},\\
2   & d=\sqrt{2}\|s_0-s_2\|=2.69\sqrt{E_s},\\
0   & otherwise.
\end{array}\right.
\end{equation} 
An analytical expression of the distance spectrum upper bound for the synthetic good channel with the standard transform is  
\begin{eqnarray}
&P_{e}\leq2Q\left(1.66\sqrt{{SNR}/{2}}\right)+2Q\left(2.69\sqrt{{SNR}/{2}}\right).&
\label{eqn_ub}
\end{eqnarray} 
As we can see from (Fig.5 in \cite{arxivSinan}), the analytic result (\ref{eqn_ub}) is a tight upper bound for the symbol error probability. 

In this paper, we first focus on designing polarizing transforms which increase the minimum distance for $q$-ary PSK signal sets. 
To design a table of transform with the type of $f(u_1,u_2)=u_1\oplus \pi(u_2)$, we define a simple procedure as follows: 
\begin{enumerate}[(p.i)]
\item Each row has only 1 candidate of $u_2$ for $u_1=0$,  
\item Each column has only 1 candidate of $u_2$ for $u_1=0$,
\item Place all candidates of $u_2$ as far as from each others for $u_1=0$,
\item Fill the empty cells by $q$ candidates of $u_2$ that are placed in the $k$th cyclic right-shift cell for $u_1=k$, where $k=1,\dots,q-1$.   
\end{enumerate} 
Hence, the completed table provides the polarizing transform with the type of $f(u_1,u_2)=u_1\oplus \pi(u_2)$.
Here, (p.i) and (p.ii) are constraints of the procedure to guarantee $(x_1,x_2)$ that can take all possible $q$-ary pairs. 
To achieve an increased minimum distance, (p.iii) can be done by a computer search. Then, (p.iv) is to complete the definition of the polarizing transform.

\begin{mylemma}
The distance is conserved as follows:
%\begin{multline}
$$\sum_{u_2'}{(\|s_{f(u_1,u_2)}-s_{f(u_1,u_2')}\|^2+\|s_{u_2}-s_{u_2'}\|^2)}
=2\sum_{k=1}^{q-1}\|s_k-s_0\|^2$$
%\nonumber
%\end{multline}
for $q$-ary PSK signal set by using a polarizing transform $f$.
\end{mylemma}
\begin{myproof}
It is obtained by the following steps:
\begin{itemize}
\item PSK is a signal set that is matched to a group \cite{loeliger_signal_1991} that $\|s_{l+k}-s_l\|=\|s_k-s_0\|$.
\item $\sum_{u_2'}{\|s_{u_2}-s_{u_2'}\|^2}=\sum_{k=1}^{q-1}\|s_k-s_0\|^2.$
\item For a fixed $u_1$, $u_2 \rightarrow f(u_1,u_2)$ is an invertible function of $u_2$. Hence, $\sum_{u_2'}{\|s_{f(u_1,u_2)}-s_{f(u_1,u_2')}\|^2}=\sum_{k=1}^{q-1}\|s_k-s_0\|^2.$
\end{itemize}
\end{myproof}

\begin{mylemma}
The minimum distance of a $q$-ary synthetic good channel is lower bounded by $$d_{min}\leq\sqrt{\frac{2}{q-1}\sum_{k=1}^{q-1}\|s_k-s_0\|^2},$$ where $s_k \in S$, and $S$ is the $q$-ary PSK signal set.
\end{mylemma}
\begin{myproof}
The distance is conserved which is seen in the previous result, and it is easy to see that 
$$\min \{d^2\}\leq \mathcal{D}/(q-1),$$
where $$\mathcal{D}=2\sum_{k=1}^{q-1}\|s_k-s_0\|^2.$$ By this way, the proof of $d_{min}\leq\sqrt{\mathcal{D}/(q-1)}$ is obtained.
\end{myproof}

\begin{myproposition}\label{theo3}
For $a> 0$, $b>0$ and $a\neq b$,
$$2Q\left(\sqrt{\frac{a^2+b^2}{2}}\right) < Q\left(a\right)+Q\left(b\right).$$ 
\label{ww}
\end{myproposition}
The proof of the proposition is provided in Appendix in \cite{arxivSinan}.

\begin{mydefinition}[Equidistant Polarizing Transforms]
For a given $q$-ary signal set the polarizing transforms with the distance spectrum $N(d_{min})=q-1$ are the Equidistant polarizing transforms.
\end{mydefinition}

\begin{mytheorem}
An equidistant transform for $q$-ary signal set has the distance spectrum upper bound as follows:
$$P_e\leq (q-1)Q(d_{min}/(2\sigma)),$$
where the minimum distance $$d_{min}=\sqrt{\frac{2}{q-1}\sum_{k=1}^{q-1}\|s_k-s_0\|^2}.$$ This is the best achievable distance spectrum bound. 
\label{www}
\end{mytheorem}

\begin{myproof}
The proof is obtained by Lemma 1, Lemma 2 and Proposition 1.
\end{myproof}

Here, we provide some examples for the better distance characteristics.

\begin{myexample}[Equidistant Polarizing Transforms for $q=5$]
We applied the procedure for 5-PSK signal set to design polarizing transforms with the type of $f(u_1,u_2)=u_1\oplus \pi(u_2)$ for $q=5$. 
Hence, the following tables are provided for $q=5$ and 5-PSK signal set. 

$$
%\hspace{-0.1cm}
\begin{tikzpicture}[scale=0.85, every node/.style={scale=0.85}]
\draw  (-1.5,0)  -- (1.5,0) ;%+++++
\draw  (-1.0,-0.5)  -- (1.5,-0.5) ;%+++++
\draw  (-1.0,-1.0)  -- (1.5,-1.0) ;%+++++
\draw  (-1.0,-1.5)  -- (1.5,-1.5) ;%+++++
\draw  (-1.0,-2.0)  -- (1.5,-2.0) ;%+++++
\draw  (-1.0,-2.5)  -- (1.5,-2.5) ;%+++++
\draw  (-1.0, 0.5)  -- (-1.0,-2.5) ;%+++++
\draw  (-0.5, 0.0)  -- (-0.5,-2.5) ;%+++++
\draw  (-0.0, 0.0)  -- (-0.0,-2.5) ;%+++++
\draw  (0.5, 0.0)  -- (0.5,-2.5) ;%+++++
\draw  (1.0, 0.0)  -- (1.0,-2.5) ;%+++++
\draw  (1.5, 0.0)  -- (1.5,-2.5) ;%+++++
\draw  (-1.5, 0.5)  -- (-1.0,0.0) ;%+++++
\draw (-1.2,0.5) node[] {$x_1$};
\draw (-0.75,0) node[above] {$0$};
\draw (-0.25,0) node[above] {$1$};
\draw ( 0.25,0) node[above] {$2$};
\draw ( 0.75,0) node[above] {$3$};
\draw ( 1.25,0) node[above] {$4$};
\draw (-1.5,-0.0) node[above] {$x_2$};
\draw (-1.25,-0.5) node[above] {$0$};
\draw (-1.25,-1.0) node[above] {$1$};
\draw (-1.25,-1.5) node[above] {$2$};
\draw (-1.25,-2.0) node[above] {$3$};
\draw (-1.25,-2.5) node[above] {$4$};
%\draw[fill=gray!50] (x-0.25,y) rectangle (x+0.25,y+0.5)
%\draw[fill=gray!40] (x-0.25,y) rectangle (x+0.25,y+0.5)
%\draw[fill=gray!30] (x-0.25,y) rectangle (x+0.25,y+0.5)
%\draw[fill=gray!20] (x-0.25,y) rectangle (x+0.25,y+0.5)
%\draw[fill=gray!10] (x-0.25,y) rectangle (x+0.25,y+0.5)
\draw[fill=gray!50] (-1.25+0.5-0.25,-0.5) rectangle (-1.25+0.5+0.25,-0.5+0.5) (-1.25+0.5,-0.5) node[above] {$00$};
\draw (-1.25+0.5,-1.0) node[above] {$31$};
\draw (-1.25+0.5,-1.5) node[above] {$12$};
\draw (-1.25+0.5,-2.0) node[above] {$43$};
\draw (-1.25+0.5,-2.5) node[above] {$24$};
\draw (-1.25+1.0,-0.5) node[above] {$10$};
\draw (-1.25+1.0,-1.0) node[above] {$41$};
\draw (-1.25+1.0,-1.5) node[above] {$22$};
\draw[fill=gray!50] (-1+0.5,-0.5-1.5) rectangle (-0.5+0.5,0-1.5) (-1.25+1.0,-2.0) node[above] {$03$};
\draw (-1.25+1.0,-2.5) node[above] {$34$};
\draw (-1.25+1.5,-0.5) node[above] {$20$};
\draw[fill=gray!50] (-1+1,-0.5-0.5) rectangle (-0.5+1,0-0.5) (-1.25+1.5,-1.0) node[above] {$01$};
\draw (-1.25+1.5,-1.5) node[above] {$32$};
\draw (-1.25+1.5,-2.0) node[above] {$13$};
\draw (-1.25+1.5,-2.5) node[above] {$44$};
\draw (-1.25+2.0,-0.5) node[above] {$30$};
\draw (-1.25+2.0,-1.0) node[above] {$11$};
\draw (-1.25+2.0,-1.5) node[above] {$42$};
\draw (-1.25+2.0,-2.0) node[above] {$23$};
\draw[fill=gray!50] (-1+1.5,-0.5-2.0) rectangle (-0.5+1.5,0-2.0)  (-1.25+2.0,-2.5) node[above] {$04$};
\draw (-1.25+2.5,-0.5) node[above] {$40$};
\draw (-1.25+2.5,-1.0) node[above] {$21$};
\draw[fill=gray!50] (-1+2,-0.5-1.0) rectangle (-0.5+2,0-1.0) (-1.25+2.5,-1.5) node[above] {$02$};
\draw (-1.25+2.5,-2.0) node[above] {$33$};
\draw (-1.25+2.5,-2.5) node[above] {$14$};
\draw (0.25,-2.5) node[below] {$\pi_1=\left(\begin{smallmatrix} 0& 1& 2& 3& 4\\ 0&2&4&1&3 \end{smallmatrix}\right)$};
\end{tikzpicture}
\begin{tikzpicture}[scale=0.85, every node/.style={scale=0.85}]
\draw  (-1.5,0)  -- (1.5,0) ;%+++++
\draw  (-1.0,-0.5)  -- (1.5,-0.5) ;%+++++
\draw  (-1.0,-1.0)  -- (1.5,-1.0) ;%+++++
\draw  (-1.0,-1.5)  -- (1.5,-1.5) ;%+++++
\draw  (-1.0,-2.0)  -- (1.5,-2.0) ;%+++++
\draw  (-1.0,-2.5)  -- (1.5,-2.5) ;%+++++
\draw  (-1.0, 0.5)  -- (-1.0,-2.5) ;%+++++
\draw  (-0.5, 0.0)  -- (-0.5,-2.5) ;%+++++
\draw  (-0.0, 0.0)  -- (-0.0,-2.5) ;%+++++
\draw  (0.5, 0.0)  -- (0.5,-2.5) ;%+++++
\draw  (1.0, 0.0)  -- (1.0,-2.5) ;%+++++
\draw  (1.5, 0.0)  -- (1.5,-2.5) ;%+++++
\draw  (-1.5, 0.5)  -- (-1.0,0.0) ;%+++++
\draw (-1.2,0.5) node[] {$x_1$};
\draw (-0.75,0) node[above] {$0$};
\draw (-0.25,0) node[above] {$1$};
\draw ( 0.25,0) node[above] {$2$};
\draw ( 0.75,0) node[above] {$3$};
\draw ( 1.25,0) node[above] {$4$};
\draw (-1.5,-0.0) node[above] {$x_2$};
\draw (-1.25,-0.5) node[above] {$0$};
\draw (-1.25,-1.0) node[above] {$1$};
\draw (-1.25,-1.5) node[above] {$2$};
\draw (-1.25,-2.0) node[above] {$3$};
\draw (-1.25,-2.5) node[above] {$4$};
\draw[fill=gray!50] (-1.25+0.5-0.25,-0.5) rectangle (-1.25+0.5+0.25,-0.5+0.5) (-1.25+0.5,-0.5) node[above] {$00$};
\draw (-1.25+0.5,-1.0) node[above] {$21$};
\draw (-1.25+0.5,-1.5) node[above] {$42$};
\draw (-1.25+0.5,-2.0) node[above] {$13$};
\draw (-1.25+0.5,-2.5) node[above] {$34$};
\draw (-1.25+1.0,-0.5) node[above] {$10$};
\draw (-1.25+1.0,-1.0) node[above] {$31$};
\draw[fill=gray!50] (-1.25+1.0-0.25,-1.5) rectangle (-1.25+1.0+0.25,-1.5+0.5) (-1.25+1.0,-1.5) node[above] {$02$};
\draw (-1.25+1.0,-2.0) node[above] {$23$};
\draw (-1.25+1.0,-2.5) node[above] {$44$};
\draw (-1.25+1.5,-0.5) node[above] {$20$};
\draw (-1.25+1.5,-1.0) node[above] {$41$};
\draw (-1.25+1.5,-1.5) node[above] {$12$};
\draw (-1.25+1.5,-2.0) node[above] {$33$};
\draw[fill=gray!50] (-1.25+1.5-0.25,-2.5) rectangle (-1.25+1.5+0.25,-2.5+0.5) (-1.25+1.5,-2.5) node[above] {$04$};
\draw (-1.25+2.0,-0.5) node[above] {$30$};
\draw[fill=gray!50] (-1.25+2.0-0.25,-1.0) rectangle (-1.25+2.0+0.25,-1.0+0.5) (-1.25+2.0,-1.0) node[above] {$01$};
\draw (-1.25+2.0,-1.5) node[above] {$22$};
\draw (-1.25+2.0,-2.0) node[above] {$43$};
\draw (-1.25+2.0,-2.5) node[above] {$14$};
\draw (-1.25+2.5,-0.5) node[above] {$40$};
\draw (-1.25+2.5,-1.0) node[above] {$11$};
\draw (-1.25+2.5,-1.5) node[above] {$32$};
\draw[fill=gray!50] (-1.25+2.5-0.25,-2.0) rectangle (-1.25+2.5+0.25,-2.0+0.5) (-1.25+2.5,-2.0) node[above] {$03$};
\draw (-1.25+2.5,-2.5) node[above] {$24$};
\draw (0.25,-2.5) node[below] {$\pi_2=\left(\begin{smallmatrix} 0& 1& 2& 3& 4\\ 0&3&1&4&2 \end{smallmatrix}\right)$};
\end{tikzpicture}$$

We investigate the distance properties for the polarizing transforms $f(u_1,u_2)=u_1\oplus \pi_i(u_2)$ for $i=1,2$, where $\pi_1=\left(\begin{smallmatrix} 0& 1& 2& 3& 4\\ 0&2&4&1&3 \end{smallmatrix}\right)$ and $\pi_2=\left(\begin{smallmatrix} 0& 1& 2& 3& 4\\ 0&3&1&4&2 \end{smallmatrix}\right)$ by using the tables that the distance properties are the same for $\pi_1$ and $\pi_2$.  Hence, the minimum distance is 
\begin{equation}
d_{\min}=\sqrt{\|s_0-s_1\|^2+\|s_0-s_2\|^2}=2.24\sqrt{E_s}
\label{dmin_eq}
\end{equation}
for 5-PSK signal set, and the distance spectrum is as follows:
\begin{equation}
N(d)=\left\{
\begin{array}{ll}
4   & d=2.24\sqrt{E_s},\\
0   & otherwise.
\end{array}\right.
\end{equation}
for the polarizing transforms $f(u_1,u_2)=u_1\oplus \pi_i(u_2)$ for $i=1,2$.
It is clear to see that these polarizing transforms are equidistant (i.e. $N(d_{min})=q-1$) for 5-PSK signal set, and the minimum distance (\ref{dmin_eq}) of the equidistant transforms is larger than the minimum distance (\ref{dmin_st}) of the standard transform. The distance spectrum upper bound is given for the equidistant transform as follows: 
\begin{eqnarray}
&P_{e}\leq4Q\left(2.24\sqrt{{SNR}/{2}}\right).&
\end{eqnarray} 

Theorem \ref{www} shows that the distance spectrum upper bound is minimized by the help of an equidistant transform for the $q$-ary signal set. 
As an analytic result for $q=5$, the standard transform provides an upper bound $2Q\left(1.66\sqrt{{SNR}/{2}}\right)+2Q\left(2.69\sqrt{{SNR}/{2}}\right)$ that is higher than $4Q\left(2.24\sqrt{{SNR}/{2}}\right)$ which is provided by the equidistant transform. This comparison can be verified by the use of Proposition 1. 
Then, we can say that the upper bound of the error performance is improved for the synthetic good channel by using equidistant transforms for a given signal set. To support the claim, simulation results show that 2dB improvement in signal to noise ratio at $10^{-3}$ symbol error rate is obtained for the synthetic good channel for $q=5$ and 5-PSK signal set by using the equidistant transform $f(u_1,u_2)=u_1\oplus \pi_1(u_2)$, where $\pi_1=\left(\begin{smallmatrix} 0& 1& 2& 3& 4\\ 0&2&4&1&3 \end{smallmatrix}\right)$.   

Reed-Solomon matrix as defined in \cite{mori_source_2014} is a mapping of this type with $f(u_1,u_2)=u_1\oplus \gamma u_2$, where $\oplus$ denotes addition mod-$q$ and $\gamma$ is a prime number. 
Abbe et al. investigated Reed-Solomon matrices in \cite{abbe_entropies_2015}. We notice that our proposed transform is equivalent to the Reed-Solomon matrix in \cite{abbe_entropies_2015}.

We follow the same way to investigate the distance properties of the synthetic bad channel by using the table. 
The analysis of the synthetic bad channel shows that the upper bounds are (almost) the same for any polarizing transform.

As such, the minimum distances of the synthetic bad channel are the same,  
\begin{eqnarray}
d_{\min}&=&\|s_i-s_{i+1}\|=1.176\sqrt{E_s},
\end{eqnarray} 
for the standard transform and the equidistant transform for $q$-ary PSK signal set.

The distance spectrum of the standard transform is 
\begin{equation}
N(d)=\left\{
\begin{array}{ll}
4   & d=\|s_0-s_1\|                                       =1.176\sqrt{E_s},\\
2   & d=\sqrt{2}\|s_0-s_1\|                            =1.663\sqrt{E_s},\\
4   & d=\|s_0-s_2\|                                       =1.902\sqrt{E_s},\\
8   & d=\sqrt{\|s_0-s_1\|^2+\|s_0-s_2\|^2}   =2.236\sqrt{E_s},\\
2   & d=\sqrt{2}\|s_0-s_2\|                           =2.690\sqrt{E_s},\\
0   & otherwise.
\end{array}\right.
\end{equation}
for $q=5$ and 5-PSK signal set.

The distance spectrum of the equidistant transform is 
\begin{equation}
N(d)=\left\{
\begin{array}{ll}
4   & d=\|s_0-s_1\|                                       =1.176\sqrt{E_s},\\
4   & d=\sqrt{2}\|s_0-s_1\|                            =1.663\sqrt{E_s},\\
4   & d=\|s_0-s_2\|                                       =1.902\sqrt{E_s},\\
4   & d=\sqrt{\|s_0-s_1\|^2+\|s_0-s_2\|^2}   =2.236\sqrt{E_s},\\
4   & d=\sqrt{2}\|s_0-s_2\|                           =2.690\sqrt{E_s},\\
0   & otherwise.
\end{array}\right.
\end{equation}
for $q=5$ and 5-PSK signal set.
The difference between the upper bounds of the synthetic bad channel is insignificant for the standard transform and the equidistant transform.

\end{myexample}

Here, one of the main results in this work is: \textsl{the equidistant transforms for a given $q$-ary signal set provide superior synthetic good channel and (almost) the same synthetic bad channel, and hence, the performance of the error correction capability is improved for the block length $N\geq2$.} 

As an alternative example, all possible polarizing transforms are always equidistant for 3-PSK signal set. 

Unfortunately, we can not find equidistant polarizing transforms for q-ary PSK signal sets for $q=4$ and $q=8$.
Alternatively, we find polarizing transforms to improve distance characteristics for 4-PSK and 8-PSK signal sets by the use of permutations $\pi=\left(\begin{smallmatrix} 0& 1& 2& 3\\ 0&2&1&3 \end{smallmatrix}\right)$ and $\pi=\left(\begin{smallmatrix} 0& 1& 2& 3& 4& 5& 6& 7 \\ 0&3&6&1&4&7&2&5 \end{smallmatrix}\right)$, respectively. It can be shown that distance spectrums are $N(d)=\{2,1\}$ and $N(d)=\{6,1\}$ for the case $q=4$ and $q=8$. The minimum distance is improved to $d_{min}=2\sqrt{E_s}$ for the case of $q=8$. 

\begin{myexample}[Almost-Equidistant Transform for $q=8$]
We applied the procedure for $q=8$ and 8-PSK signal set to design a polarizing transform. There is not exist an equidistant transform of type $f(u_1,u_2)=u_1\oplus \pi(u_2)$.
Thanks to the following geometric property of 8-PSK signal set, we can design \textsl{almost}-equidistant transform of type $f(u_1,u_2)=u_1\oplus \pi(u_2)$, where $\pi=\left(\begin{smallmatrix} 0& 1& 2& 3& 4& 5& 6& 7\\ 0&3&6&1&4&7&2&5 \end{smallmatrix}\right)$  by using the procedure for $q=8$.
$${\|s_0-s_1\|^2+\|s_0-s_3\|^2}={2}\|s_0-s_2\|^2=\|s_0-s_4\|^2$$
This geometric property is depicted in Fig.~\ref{Fig_8psk}.

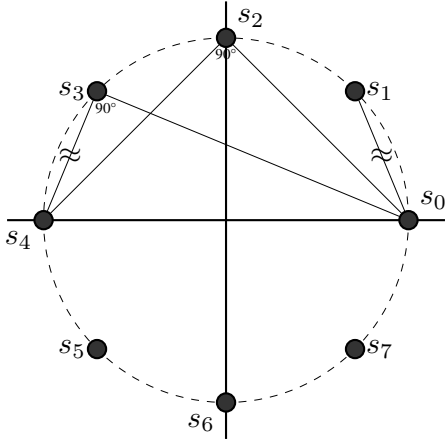
\begin{figure}[hp]
\centering
\begin{tikzpicture}[thick,scale=1.5*1.1*1.4*3.0*0.35, every node/.style={scale=1.1*1.1*1.5*1.25*0.5}]
\draw  (-1.2,0)  -- (1.2,0) ;
\draw  (0,-1.2)  -- (0,1.2) ;
\begin{scope}[very thin]
\draw[<->]  (1,0)  -- (1/1.414213562373095,1/1.414213562373095) ;
\draw[<->]  (1,0)  -- (-1/1.414213562373095,1/1.414213562373095) ;
\draw[<->]  (-1,0)  -- (-1/1.414213562373095,1/1.414213562373095) ;
\draw[<->]  (1,0)  -- (0,1) ;
\draw[<->]  (-1,0)  -- (0,1) ;
%\draw  (0.333,0.25) node[above] {$\begin{smallmatrix}\sqrt{E_s}\end{smallmatrix}$};
\end{scope}
\begin{scope}[very thin,dashed]
\draw (0,0) circle (1.0cm);
\end{scope}
\draw[fill=gray!160!white] (1,0) circle (0.05cm) node[above right] {$s_0$} node[below right] {$$};
\draw[fill=gray!160!white] (1/1.414213562373095,1/1.414213562373095) circle (0.05cm) node[right] {$s_1$} node[ right] {$$};
\draw (0.5+0.5/1.414213562373095,0.5/1.414213562373095)  node[] {$\approx$};
\draw (-0.5-0.5/1.414213562373095,0.5/1.414213562373095)  node[] {$\approx$};
\draw[fill=gray!160!white] (0,1) circle (0.05cm) node[above right] {$s_2$};
\draw (0.00025,0.9995)  node[below] {\begin{tiny}90\textdegree \end{tiny}};
\draw[fill=gray!160!white] (-1/1.414213562373095,1/1.414213562373095) circle (0.05cm) node[left] {$s_3$};
\draw (-1/1.414213562373095+0.05,1/1.414213562373095) node[below] {\begin{tiny}90\textdegree \end{tiny}};
\draw[fill=gray!160!white] (-1.0,0.0) circle (0.05cm) node[below left] {$s_4$} node[above left] {$$};
\draw[fill=gray!160!white] (-1/1.414213562373095,-1/1.414213562373095) circle (0.05cm) node[left] {$s_5$} node[ right] {$$};
\draw[fill=gray!160!white] (0,-1) circle (0.05cm) node[below left] {$s_6$} node[below left] {$$};
\draw[fill=gray!160!white] (1/1.414213562373095,-1/1.414213562373095) circle (0.05cm) node[right] {$s_7$} node[ right] {$$};
\end{tikzpicture}
%\begin{tikzpicture}[thick,scale=1.5, every node/.style={scale=1.5}]
%\draw  (-1.2,0)  -- (1.2,0) ;
%\draw  (0,-1.2)  -- (0,1.2) ;
%\begin{scope}[very thin]
%\draw[<->]  (0,0)  -- (0.866,0.5) ;
%\draw  (0.333,0.25) node[above] {$\begin{smallmatrix}\sqrt{E_s}\end{smallmatrix}$};
%\end{scope}
%\begin{scope}[very thin,dashed]
%\draw (0,0) circle (1.0cm);
%\end{scope}
%\draw[fill=gray!160!white] (1,0) circle (0.05cm) node[above right] {$s_0$};
%\draw[fill=gray!160!white] (0.0,1.0) circle (0.05cm) node[above] {$s_1$};
%\draw[fill=gray!160!white] (-1.0,0.0) circle (0.05cm) node[below] {$s_2$};
%\draw[fill=gray!160!white] (0.0,-1.0) circle (0.05cm) node[below] {$s_3$};
%\end{tikzpicture}
\caption{The geometric property of $8$-PSK signal set for almost-equidistant transform.}
\label{Fig_8psk}
\end{figure}

The minimum distance of the almost-equidistant transform is $d_{min}=2\sqrt{E_s}$.
The distance spectrum of the almost-equidistant transform is
\begin{equation}
N(d)=\left\{
\begin{array}{ll}
6   & d=2\sqrt{E_s},\\
1   & d=\sqrt{\|s_0-s_4\|^2+\|s_0-s_4\|^2}=2.83\sqrt{E_s},\\
0   & otherwise.
\end{array}\right.
\end{equation}
for $q=8$ and 8-PSK signal set.
Hence, the distance spectrum upper bound of the almost-equidistant transform is
\begin{eqnarray}
&P_{e}\leq6Q\left(2\sqrt{\frac{SNR}{2}}\right)+Q\left(2.83\sqrt{\frac{SNR}{2}}\right).&
\end{eqnarray} 
If an equidistant transform existed for 8-PSK signal set, the minimum distance would be $d_{min}=2.14\sqrt{E_s}$, and the upper bound would be  
\begin{eqnarray}
&P_{e}\leq7Q\left(2.14\sqrt{\frac{SNR}{2}}\right).&
\end{eqnarray} 
The proposed transform has also (almost) the same with the equidistant upper bound for 8-PSK signal set that can be seen in the Appendix in \cite{arxivSinan}.
\end{myexample}

\subsection{Experiments on the almost-equidistant transform}\label{S_speed_polarization_exp}
\begin{figure}[!t]
\centering
\begin{tikzpicture}[thick,scale=1.10*0.3*1.0, every node/.style={scale=1.10*0.45*2*1.0}]
\draw[->]  (-1,0)  -- (-0.2,0) ;
\draw  (0.7,0)  -- (1+1-0.5,0) ;
\draw[->]  (0.25,-2)  -- (0.25,-1.2) ;
\draw  (-1,-2)  -- (1+1-0.5,-2) ;
\draw (-0.2,-1.2)  -- (0.7,-1.2);
\draw (-0.2,0.5)  -- (0.7,0.5);
\draw (-0.2,-1.2)  -- (-0.2,0.5);
\draw (0.7,-1.2)  -- (0.7,0.5);
\begin{scope}[thick]
\draw (0.25,-0.35) node[] {$+$};
\end{scope}
\draw (-1,0) node[left] {$u_1$};
\draw (-1,-2) node[left] {$u_2$};
\draw (1+2-1.5,0)  -- (-1+4.5,0);
\draw (1+2-1.5,-2)  -- (-1+4.5,0-4);
\draw[->]  (-1,0-4)  -- (-0.2,0-4) ;
\draw  (0.7,0-4)  -- (1+1-0.5,0-4) ;
\draw[->]  (0.25,-2-4)  -- (0.25,-1.2-4) ;
\draw  (-1,-2-4)  -- (1+1-0.5,-2-4) ;
\draw (-0.2,-1.2-4)  -- (0.7,-1.2-4);
\draw (-0.2,0.5-4)  -- (0.7,0.5-4);
\draw (-0.2,-1.2-4)  -- (-0.2,0.5-4);
\draw (0.7,-1.2-4)  -- (0.7,0.5-4);
\begin{scope}[thick]
\draw (0.25,-0.35-4) node[] {$+$};
\end{scope}
\draw (-1,0-4) node[left] {$u_3$};
\draw (-1,-2-4) node[left] {$u_4$};
\draw (1+2-1.5,0-4)   -- (-1+4.5,-2);
\draw (1+2-1.5,-2-4)  -- (-1+4.5,-2-4);
\draw[->]  (-1+4.5,0)  -- (-0.2+4.5,0) ;
\draw  (0.7+4.5,0)  -- (1+1-0.5+4.5,0) ;
\draw[->]  (0.25+4.5,-2)  -- (0.25+4.5,-1.2) ;
\draw  (-1+4.5,-2)  -- (1+1-0.5+4.5,-2) ;
\draw (-0.2+4.5,-1.2)  -- (0.7+4.5,-1.2);
\draw (-0.2+4.5,0.5)  -- (0.7+4.5,0.5);
\draw (-0.2+4.5,-1.2)  -- (-0.2+4.5,0.5);
\draw (0.7+4.5,-1.2)  -- (0.7+4.5,0.5);
\begin{scope}[thick]
\draw (0.25+4.5,-0.35) node[] {$+$};
\end{scope}
%\draw (1+2-1.5+4.5,0-0.03) node[right] {$x_1$};
%\draw (1+2-1.5+4.5,-2-0.03) node[right] {$x_2$};
%
\draw[->]  (-1+4.5,0-4)  -- (-0.2+4.5,0-4) ;
\draw  (0.7+4.5,0-4)  -- (1+1-0.5+4.5,0-4) ;
\draw[->]  (0.25+4.5,-2-4)  -- (0.25+4.5,-1.2-4) ;
\draw  (-1+4.5,-2-4)  -- (1+1-0.5+4.5,-2-4) ;
\draw (-0.2+4.5,-1.2-4)  -- (0.7+4.5,-1.2-4);
\draw (-0.2+4.5,0.5-4)  -- (0.7+4.5,0.5-4);
\draw (-0.2+4.5,-1.2-4)  -- (-0.2+4.5,0.5-4);
\draw (0.7+4.5,-1.2-4)  -- (0.7+4.5,0.5-4);
\begin{scope}[thick]
\draw (0.25+4.5,-0.35-4) node[] {$+$};
\end{scope}
%\draw (1+2-1.5+4.5,0-0.03-4) node[right] {$x_3$};
%\draw (1+2-1.5+4.5,-2-0.03-4) node[right] {$x_4$};
%%%
\draw[->]  (-1,0-8)  -- (-0.2,0-8) ;
\draw  (0.7,0-8)  -- (1+1-0.5,0-8) ;
\draw[->]  (0.25,-2-8)  -- (0.25,-1.2-8) ;
\draw  (-1,-2-8)  -- (1+1-0.5,-2-8) ;
\draw (-0.2,-1.2-8)  -- (0.7,-1.2-8);
\draw (-0.2,0.5-8)  -- (0.7,0.5-8);
\draw (-0.2,-1.2-8)  -- (-0.2,0.5-8);
\draw (0.7,-1.2-8)  -- (0.7,0.5-8);
\begin{scope}[thick]
\draw (0.25,-0.35-8) node[] {$+$};
\end{scope}
\draw (-1,0-8) node[left] {$u_5$};
\draw (-1,-2-8) node[left] {$u_6$};
\draw (1+2-1.5,0-8)  -- (-1+4.5,0-8);
\draw (1+2-1.5,-2-8)  -- (-1+4.5,0-4-8);
\draw[->]  (-1,0-4-8)  -- (-0.2,0-4-8) ;
\draw  (0.7,0-4-8)  -- (1+1-0.5,0-4-8) ;
\draw[->]  (0.25,-2-4-8)  -- (0.25,-1.2-4-8) ;
\draw  (-1,-2-4-8)  -- (1+1-0.5,-2-4-8) ;
\draw (-0.2,-1.2-4-8)  -- (0.7,-1.2-4-8);
\draw (-0.2,0.5-4-8)  -- (0.7,0.5-4-8);
\draw (-0.2,-1.2-4-8)  -- (-0.2,0.5-4-8);
\draw (0.7,-1.2-4-8)  -- (0.7,0.5-4-8);
\begin{scope}[thick]
\draw (0.25,-0.35-4-8) node[] {$+$};
\end{scope}
\draw (-1,0-4-8) node[left] {$u_7$};
\draw (-1,-2-4-8) node[left] {$u_8$};
\draw (1+2-1.5,0-4-8)   -- (-1+4.5,-2-8);
\draw (1+2-1.5,-2-4-8)  -- (-1+4.5,-2-4-8);
\draw[->]  (-1+4.5,0-8)  -- (-0.2+4.5,0-8) ;
\draw  (0.7+4.5,0-8)  -- (1+1-0.5+4.5,0-8) ;
\draw[->]  (0.25+4.5,-2-8)  -- (0.25+4.5,-1.2-8) ;
\draw  (-1+4.5,-2-8)  -- (1+1-0.5+4.5,-2-8) ;
\draw (-0.2+4.5,-1.2-8)  -- (0.7+4.5,-1.2-8);
\draw (-0.2+4.5,0.5-8)  -- (0.7+4.5,0.5-8);
\draw (-0.2+4.5,-1.2-8)  -- (-0.2+4.5,0.5-8);
\draw (0.7+4.5,-1.2-8)  -- (0.7+4.5,0.5-8);
\begin{scope}[thick]
\draw (0.25+4.5,-0.35-8) node[] {$+$};
\end{scope}
%\draw (1+2-1.5+4.5,0-0.03-8) node[right] {$x_1$};
%\draw (1+2-1.5+4.5,-2-0.03-8) node[right] {$x_2$};
%
\draw[->]  (-1+4.5,0-4-8)  -- (-0.2+4.5,0-4-8) ;
\draw  (0.7+4.5,0-4-8)  -- (1+1-0.5+4.5,0-4-8) ;
\draw[->]  (0.25+4.5,-2-4-8)  -- (0.25+4.5,-1.2-4-8) ;
\draw  (-1+4.5,-2-4-8)  -- (1+1-0.5+4.5,-2-4-8) ;
\draw (-0.2+4.5,-1.2-4-8)  -- (0.7+4.5,-1.2-4-8);
\draw (-0.2+4.5,0.5-4-8)  -- (0.7+4.5,0.5-4-8);
\draw (-0.2+4.5,-1.2-4-8)  -- (-0.2+4.5,0.5-4-8);
\draw (0.7+4.5,-1.2-4-8)  -- (0.7+4.5,0.5-4-8);
\begin{scope}[thick]
\draw (0.25+4.5,-0.35-4-8) node[] {$+$};
\end{scope}
\draw (1+2-1.5+9/2,0)  -- (-1+4.5+9/2,0);
\draw (1+2-1.5+9/2,-2)  -- (-1+4.5+9/2,0-4);

\draw (1+2-1.5+9/2,0-4)   -- (-1+4.5+9/2,-8);
\draw (1+2-1.5+9/2,-2-4)  -- (-1+4.5+9/2,-2-4-6);
\draw[->]  (-1+4.5+9/2,0)  -- (-0.2+4.5+9/2,0) ;
\draw[->]  (0.7+4.5+9/2,0)  -- (1+1-0.5+4.5+9/2,0) ;
\draw[->]  (0.25+4.5+9/2,-2)  -- (0.25+4.5+9/2,-1.2) ;
\draw[->]  (-1+4.5+9/2,-2)  -- (1+1-0.5+4.5+9/2,-2) ;
\draw (-0.2+4.5+9/2,-1.2)  -- (0.7+4.5+9/2,-1.2);
\draw (-0.2+4.5+9/2,0.5)  -- (0.7+4.5+9/2,0.5);
\draw (-0.2+4.5+9/2,-1.2)  -- (-0.2+4.5+9/2,0.5);
\draw (0.7+4.5+9/2,-1.2)  -- (0.7+4.5+9/2,0.5);
\begin{scope}[thick]
\draw (0.25+4.5+9/2,-0.35) node[] {$f$};
\end{scope}
\draw (1+2-1.5+4.5+9/2,0-0.03) node[right] {$x_1$};
\draw (1+2-1.5+4.5+9/2,-2-0.03) node[right] {$x_2$};
\draw[->]  (-1+4.5+9/2,0-4)  -- (-0.2+4.5+9/2,0-4) ;
\draw[->]  (0.7+4.5+9/2,0-4)  -- (1+1-0.5+4.5+9/2,0-4) ;
\draw[->]  (0.25+4.5+9/2,-2-4)  -- (0.25+4.5+9/2,-1.2-4) ;
\draw[->]  (-1+4.5+9/2,-2-4)  -- (1+1-0.5+4.5+9/2,-2-4) ;
\draw (-0.2+4.5+9/2,-1.2-4)  -- (0.7+4.5+9/2,-1.2-4);
\draw (-0.2+4.5+9/2,0.5-4)  -- (0.7+4.5+9/2,0.5-4);
\draw (-0.2+4.5+9/2,-1.2-4)  -- (-0.2+4.5+9/2,0.5-4);
\draw (0.7+4.5+9/2,-1.2-4)  -- (0.7+4.5+9/2,0.5-4);
\begin{scope}[thick]
\draw (0.25+4.5+9/2,-0.35-4) node[] {$f$};
\end{scope}
\draw (1+2-1.5+4.5+9/2,0-0.03-4) node[right] {$x_3$};
\draw (1+2-1.5+4.5+9/2,-2-0.03-4) node[right] {$x_4$};
\draw (1+2-1.5+9/2,0-8)  -- (-1+4.5+9/2,0-8+6);
\draw (1+2-1.5+9/2,-2-8)  -- (-1+4.5+9/2,0-4-8+6);
\draw (1+2-1.5+9/2,0-4-8)   -- (-1+4.5+9/2,-2-8);
\draw (1+2-1.5+9/2,-2-4-8)  -- (-1+4.5+9/2,-2-4-8);
\draw[->]  (-1+4.5+9/2,0-8)  -- (-0.2+4.5+9/2,0-8) ;
\draw[->]  (0.7+4.5+9/2,0-8)  -- (1+1-0.5+4.5+9/2,0-8) ;
\draw[->]  (0.25+4.5+9/2,-2-8)  -- (0.25+4.5+9/2,-1.2-8) ;
\draw[->]  (-1+4.5+9/2,-2-8)  -- (1+1-0.5+4.5+9/2,-2-8) ;
\draw (-0.2+4.5+9/2,-1.2-8)  -- (0.7+4.5+9/2,-1.2-8);
\draw (-0.2+4.5+9/2,0.5-8)  -- (0.7+4.5+9/2,0.5-8);
\draw (-0.2+4.5+9/2,-1.2-8)  -- (-0.2+4.5+9/2,0.5-8);
\draw (0.7+4.5+9/2,-1.2-8)  -- (0.7+4.5+9/2,0.5-8);
\begin{scope}[thick]
\draw (0.25+4.5+9/2,-0.35-8) node[] {$f$};
\end{scope}
\draw (1+2-1.5+4.5+9/2,0-0.03-8) node[right] {$x_5$};
\draw (1+2-1.5+4.5+9/2,-2-0.03-8) node[right] {$x_6$};
\draw[->]  (-1+4.5+9/2,0-4-8)  -- (-0.2+4.5+9/2,0-4-8) ;
\draw[->]  (0.7+4.5+9/2,0-4-8)  -- (1+1-0.5+4.5+9/2,0-4-8) ;
\draw[->]  (0.25+4.5+9/2,-2-4-8)  -- (0.25+4.5+9/2,-1.2-4-8) ;
\draw[->]  (-1+4.5+9/2,-2-4-8)  -- (1+1-0.5+4.5+9/2,-2-4-8) ;
\draw (-0.2+4.5+9/2,-1.2-4-8)  -- (0.7+4.5+9/2,-1.2-4-8);
\draw (-0.2+4.5+9/2,0.5-4-8)  -- (0.7+4.5+9/2,0.5-4-8);
\draw (-0.2+4.5+9/2,-1.2-4-8)  -- (-0.2+4.5+9/2,0.5-4-8);
\draw (0.7+4.5+9/2,-1.2-4-8)  -- (0.7+4.5+9/2,0.5-4-8);
\begin{scope}[thick]
\draw (0.25+4.5+9/2,-0.35-4-8) node[] {$f$};
\end{scope}
\draw (1+2-1.5+4.5+9/2,0-0.03-4-8) node[right] {$x_7$};
\draw (1+2-1.5+4.5+9/2,-2-0.03-4-8) node[right] {$x_8$};
\end{tikzpicture}
\caption{An encoder scheme of non-binary polar codes with $f$ at only the channel stage. Other stages have the standard transform.}
\label{Fig_chst}
\end{figure}
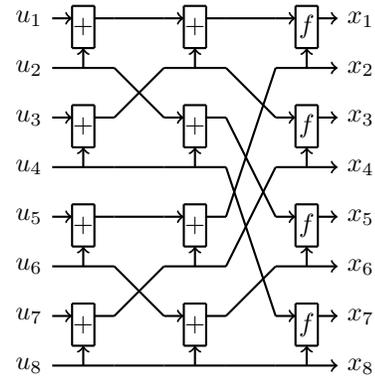

We experimentally investigate the contribution of equidistant polarization transforms that are placed at only the channel stage as shown in Fig.~\ref{Fig_chst}. 

\begin{figure}[hp]
\centering
\includegraphics[width=0.45\textwidth]{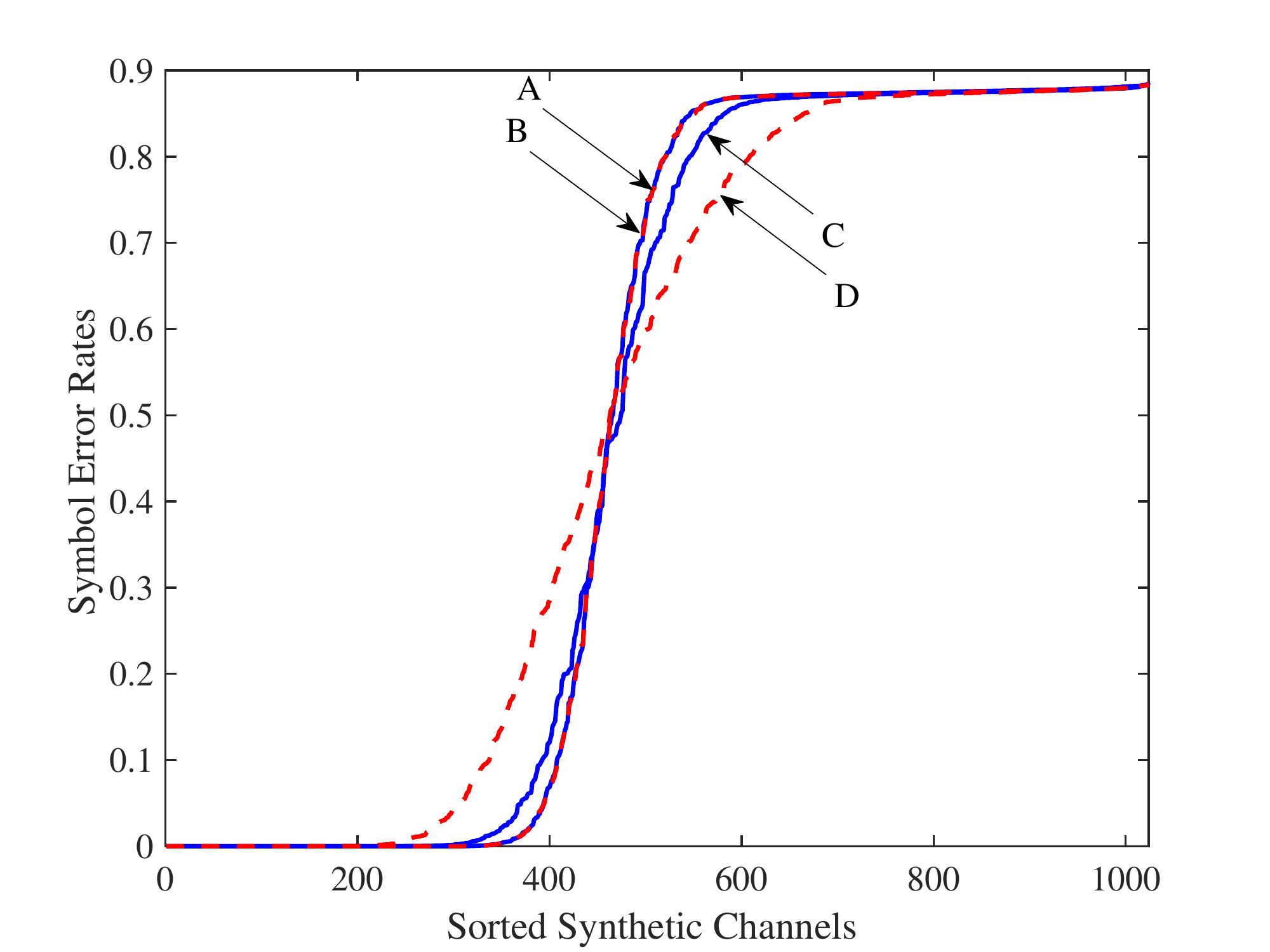}
\caption{Reliabilities of indices for $N=1024$ 8-ary polar codes. A-all stages with equidistant transform, B-channel stage with equidistant transform, others with standard,C-all stages with a non-binary transform, D-channel stage with a non-binary transform, others with standard. }
\label{fig_fin}
\end{figure}

It was depicted in Fig.~\ref{fig_fin} that contribution of any polarization transforms from the first stage to the channel state are insignificant when the equidistant transform is placed at the channel stage. This result is not true for other (not equidistant) polarization transforms. 

By similar argument, the equidistant transform that is placed at the channel stage creates an equidistant synthetic good channel, and hence, the standard transforms that are placed at all of the previous stages provide the polarization, regardless of the input alphabet size. 
This assertion is provided by the computer simulation in Fig.~\ref{fig_fin}.

\section{Simulation results}
\begin{figure}[!t]
\centering
\includegraphics[width=0.45\textwidth]{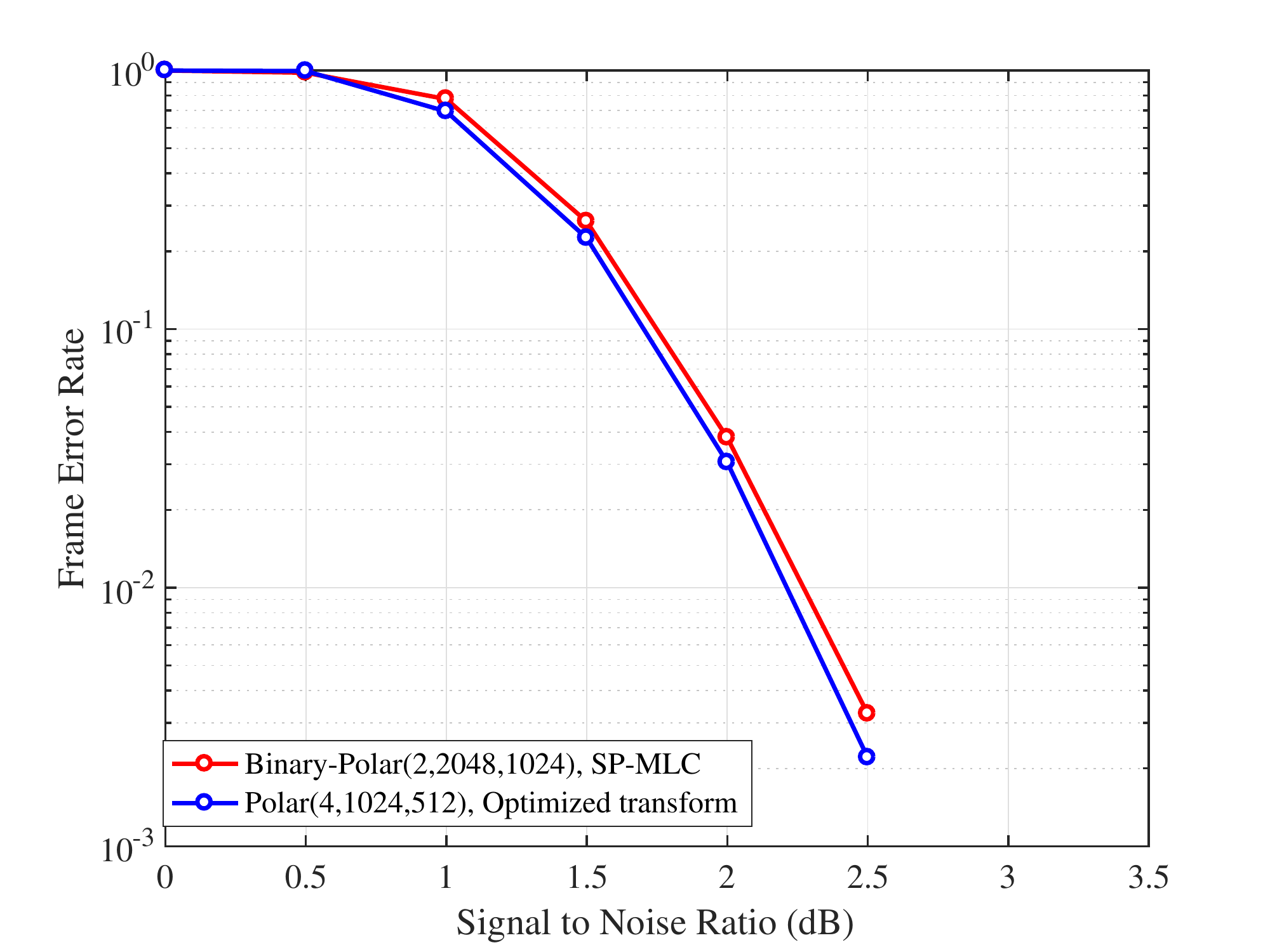}%
\caption{Frame error rates of non-binary polar code and SP-MLC binary polar code for 4-PSK signal set.}
\label{fig_fer2}
\end{figure}

The frame error rates are provided in Fig.\ref{fig_fer2} for the optimized transform for $q=4$, and 4-PSK signal set. 
Note that we compare our results with the set partitioned (SP) binary polar codes \cite{seidl_polar-coded_2013} which are state-of-the-art transmission schemes with multi-level coding (MLC) for polar codes for 4-PSK signalling.
The optimized transform outperforms state-of-the-art MLC. 

Furthermore, proposed non-binary codes in this study can also be adopted to the set partitioned MLC schemes to improve the error performance. This is out of the scope in this paper. We left this research direction to the future work.

\section{Signal-Set Design for Nonbinary Equidistance}\label{S_signalset_design}  

\begin{figure}[b!]
\centering
\begin{tikzpicture}[thick,scale=0.5*1.1*2.5*1.1*3.0*0.35, every node/.style={scale=0.9*1.4*1.25*0.5}]
\draw  (-1.2,0)  -- (1.2,0) ;
\draw  (0,-1.2)  -- (0,1.2) ;
\begin{scope}[very thin]
\draw[<->]  (0,0)  -- (0.866,0.5) ;
\draw  (0.333,0.25) node[above] {$\begin{smallmatrix}\sqrt{E_s}\end{smallmatrix}$};
\end{scope}
\begin{scope}[very thin,dashed]
\draw (0,0) circle (1.0cm);
\end{scope}
\draw[fill=gray!160!white] (1,0) circle (0.05cm) node[above right] {$s_0$} node[below] {$(1,0)$};
\begin{scope}[very thin,dashed]
\draw[fill=gray!10!white]  (0,1) circle (0.05cm);
\draw[fill=gray!10!white]  (0,-1) circle (0.05cm);
\end{scope}
\draw[<-] (1.2*1/3,1.2*0.94280904158) arc (69:87:1.2);
\draw[<-] (-1.2*1/3,-1.2*0.94280904158) arc (69+180:87+180:1.2);
\draw[fill=gray!160!white] (1/3,0.94280904158) circle (0.05cm) node[above] {$s_1$} node[ right] {$\left(\frac{1}{3},\frac{2\sqrt{2}}{3}\right)$};
\draw[fill=gray!160!white] (-1.0,0.0) circle (0.05cm) node[below left] {$s_2$} node[above] {$(-1,0)$};
\draw[fill=gray!160!white] (-1/3,-0.94280904158) circle (0.05cm) node[below] {$s_3$} node[ left] {$\left(-\frac{1}{3},-\frac{2\sqrt{2}}{3}\right)$};
\end{tikzpicture}
\caption{The new signal-set with $E_s=1$ joule/2-dimension for $q=4$ nonbinary equidistance.}
\label{Fig_6}
\end{figure}
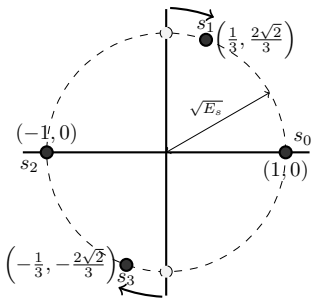

The new signal-set for $q=4$ is obtained by the changing the geometry of the 4-PSK signaling as shown in Fig.~\ref{Fig_6}. 
In this way, an equal rotation in clockwise direction was applied to the $s_1$ and $s_3$. 
The following relation is found between the new Euclidean distances, depending on the amount of rotation as a result of the rotation process. 
$$\|s_0-s_1\|=x\sqrt{E_s},$$ 
$$\|s_0-s_3\|=\sqrt{4-x^2}\sqrt{E_s}.$$
The following equation must be satisfied in order to have the equidistant characteristic of the transform given above. 
$$\sqrt{\|s_0-s_1\|^2+\|s_0-s_2\|^2}=\sqrt{\|s_0-s_3\|^2+\|s_0-s_3\|^2}$$ 
For this, $x=\frac{2}{\sqrt{3}}$ is found and the amount of rotation is determined.
As a result of these, PSK-type the new signal-set is depicted in Fig.~\ref{Fig_6}.

The distance properties are optimized by the nonbinary equidistance for the new signal-set for a fixed $u_1=2$ as follows
The minimum distance is $d_{min}=2.309\sqrt{E_s}$
The distance spectrum is $N(d_{min})=3$.

%%%%%%%
%%%%%%%

A similar way can be followed to design signal-sets in 1-dimension for nonbinary equidistance. 
It is possible to consider a PAM-type new signal-set for $q=3$ in 1-dimension for nonbinary equidistance.

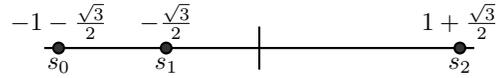
\begin{figure}[!t]
\centering
\vspace{0.01cm}
\begin{tikzpicture}[thick,scale=0.45*1.1*2.5*1.1*3.0*0.35, every node/.style={scale=1.1*1.4*1.25*0.5}]
\draw  (-2,0)  -- (2,0) ;
\draw  (0,-0.2)  -- (0,0.2) ;
\begin{scope}[very thin]
\end{scope}
\begin{scope}[very thin,dashed]
\end{scope}
\draw[fill=gray!160!white] (-1.866,0) circle (0.05cm) node[below ] {$s_0$} node[above] {$-1-\frac{\sqrt{3}}{2}$};
\begin{scope}[very thin,dashed]
\end{scope}
\draw[fill=gray!160!white] (-0.866,0.0) circle (0.05cm) node[below] {$s_1$} node[above] {$-\frac{\sqrt{3}}{2}$};
\draw[fill=gray!160!white] (1.866,0.0) circle (0.05cm) node[below] {$s_2$} node[above] {$1+\frac{\sqrt{3}}{2}$};
\end{tikzpicture}
\vspace{0.01cm}
\caption{The new signal-set for $q=3$ with $E_s=2.57$ joule/1-dimension for $q=3$ nonbinary equidistance.}
\label{Fig_pam}
\end{figure}

The new signal-set is obtained by the changing the geometry of the 3-PAM signaling as shown in Fig.~\ref{Fig_pam}. 
The following relation is found between the new Euclidean distances, depending on the amount of shift as a result of the rotation process. 
$$\|s_0-s_1\|=\alpha,\|s_1-s_2\|=\beta.$$
The following equation must be satisfied in order to have the equidistant characteristic of the transform given above. 
$$\sqrt{\|s_0-s_1\|^2+\|s_0-s_2\|^2}=\sqrt{\|s_1-s_2\|^2+\|s_1-s_2\|^2}$$ 
For this, $\beta=(1+\sqrt{3})\alpha$ is found and the amount of shift is determined.
As a result of these, the new signal-set is depicted in Fig.~\ref{Fig_pam}.

The distance properties are optimized by the nonbinary equidistance for the new signal-set as follows
The minimum distance is $d_{min}=2.415\sqrt{E_s}$
The distance spectrum is $N(d_{min})=2$.

This result provides the best distance spectrum bound for 1-dimensional signaling for $q=3$.

Simulation results for $q=4$ are shown in in \cite{arxivSinan}. 

%%%%%%%%
%%%%%%%%

\section{Conclusion}\label{S_conclusion}
We have shown how to design polarizing transforms of non-binary polar codes to improve the minimum distance. 
For a given signal set we have designed polarization transforms that reach the limit of the best known distance spectrum. 
This limit is tight for AWGN channels.
The increase in the polarization rate and even the improvement in the frame error rate are confirmed by the simulation results.

\section*{Acknowledgment}

Study was supported by T\"UB\.ITAK. I would like to thank Prof. E. Viterbo for helpful discussion on the signal set design.

\bibliography{q-ary_Ref}{}

\begin{thebibliography}{1}
\providecommand{\url}[1]{#1}
\csname url@samestyle\endcsname
\providecommand{\newblock}{\relax}
\providecommand{\bibinfo}[2]{#2}
\providecommand{\BIBentrySTDinterwordspacing}{\spaceskip=0pt\relax}
\providecommand{\BIBentryALTinterwordstretchfactor}{4}
\providecommand{\BIBentryALTinterwordspacing}{\spaceskip=\fontdimen2\font plus
\BIBentryALTinterwordstretchfactor\fontdimen3\font minus
  \fontdimen4\font\relax}
\providecommand{\BIBforeignlanguage}[2]{{%
\expandafter\ifx\csname l@#1\endcsname\relax
\typeout{** WARNING: IEEEtran.bst: No hyphenation pattern has been}%
\typeout{** loaded for the language `#1'. Using the pattern for}%
\typeout{** the default language instead.}%
\else
\language=\csname l@#1\endcsname
\fi
#2}}
\providecommand{\BIBdecl}{\relax}
\BIBdecl

\bibitem{arikan_channel_2009}
E.~Ar\i{}kan, ``Channel polarization: A method for constructing
  capacity-achieving codes for symmetric binary-input memoryless channels,''
  \emph{IEEE Trans. Inf. Theory}, vol.~55, no.~7, pp. 3051--3073, Jul. 2009.

\bibitem{arxivSinan}
S.~Kahraman, ``Equidistant polarizing transforms,'' \emph{ArXiv Prepr.
  ArXiv1708.01233}, 2017.

\bibitem{loeliger_signal_1991}
H.-A. Loeliger, ``Signal sets matched to groups,'' \emph{IEEE Trans. Inf.
  Theory}, vol.~37, no.~6, pp. 1675--1682, Nov. 1991.

\bibitem{mori_source_2014}
R.~Mori and T.~Tanaka, ``Source and channel polarization over finite fields and
  {{Reed}}-{{Solomon}} matrices,'' \emph{IEEE Trans. Inf. Theory}, vol.~60,
  no.~5, pp. 2720--2736, May 2014.

\bibitem{abbe_entropies_2015}
E.~Abbe, J.~Li, and M.~Madiman, ``Entropies of weighted sums and applications
  to polar codes,'' \emph{ArXiv Prepr. ArXiv151200135}, 2015.

\bibitem{seidl_polar-coded_2013}
M.~Seidl, A.~Schenk, C.~Stierstorfer, and J.~B. Huber, ``Polar-coded
  modulation,'' \emph{IEEE Trans. Commun.}, vol.~61, no.~10, pp. 4108--4119,
  Oct. 2013.

\end{thebibliography}
\bibliographystyle{IEEEtran}

\end{document}